\documentclass{aastex}
\usepackage{spr-astr-addons}
\usepackage{url}
\RequirePackage{flushend}
\RequirePackage{float,multirow,dcolumn,amsmath,amssymb,cuted,flushend,multicol}
\RequirePackage[colorlinks,citecolor=blue,urlcolor=blue,linkcolor=blue]{hyperref}
\RequirePackage{color}
\RequirePackage{adjustbox}
\begin{document}

\title{Analytical model on mass limit of strange stars}


\author{Sajahan Molla\altaffilmark{1}}\affil{sajahan.phy@gmail.com} 
\and 
\author{Masum Murshid\altaffilmark{2}}\affil{masummurshid2012@gmail.com} 
\and 
\author{Mehedi Kalam\altaffilmark{2}}\affil{kalam@associates.iucaa.in}
\altaffiltext{1}{Department of Physics, New Alipore College, L Block, New Alipore, Kolkata 700053, India}
\altaffiltext{2}{Department of Physics, Aliah University, IIA/27, New Town, Kolkata 700160, India}



\begin{abstract}
In this paper, we present a new kind of stellar model using the Nariai IV metric.This model can be used to study the strange/quark stars(which is our present interest, though it can also be applicable to neutron stars). We present a mass-radius region where all the regularity conditions, energy conditions, the TOV equation, and stability conditions are satisfied.  According to our model, strange stars having mass up to $1.9165M_{\odot}(=2.81 km)$ is stable. A strange star having a mass greater than $1.9165M_{\odot}$ violates the stability conditions.This model can be very useful to predict the radius of strange stars of mass greater than $1 M_{\odot}$.
\end{abstract}


\section{Introduction}
\label{intro}
The study of the compact stars drew attention to many scientists for numerous years. The compact stars are the end product of the thermonuclear powered stars. The mass of the thermonuclear powered stars decides which state of the compact star would it be. The degeneracy pressure of the constituting fermions (electrons for the white dwarf and neutrons for the neutron star) of the compact stars balances the self-gravitational force. A neutron star is the extreme state of the compact star before it turns into a black hole while the gravitational force overwhelms the degeneracy pressure. Neutron star also has several substates like hyperon star, hybrid star or strange star.  These three sub-states of the neutron star are just hypothetical states, no direct observation yet found. The strange star (made of strange quarks) and a normal neutron star (fundamentally constructed of neutrons) can be distinguished in three different ways \cite{Xu2001}, which are based on the mass-radius relation, differences of viscosity and surface condition. The neutron and strange stars have almost same radius while the mass of the star is about $1 M_{\odot}$. Whereas the low mass neutron and strange stars radius significantly differ from each other \cite{Alcock1986,Bombaci1997,Li1999,Xu2005}.  A neutron star having mass $\sim 0.2 M_{\odot}$ has a radius of $ > 15 km $, whereas a strange star with mass $\sim 0.2 M_{\odot}$ is only $ < 5 km $.  Another difference is that a neutron star has a lower mass limit of about $\sim 0.2 M_{\odot}$, whereas a strange star does not have any lower mass limit \cite{Xu2005}. 

\par In four-dimensional spacetime general relativity describes the gravitational interaction and its consequences very nicely. Einstein's theory of general relativity placed the ground of our understanding of compact stars. In 1916 Schwarzschild  \cite{Schwarzschild1916} first solved the exact solution of Einstein's field equations. He had calculated the gravitational field equation of a homogeneous incompressible fluid sphere which had finite radius. Oppenheimer, Volkoff and Tolman \cite{Oppenheimer1939,Tolman1939} in 1939 successfully derived the balancing equations of relativistic stellar structures from Einstein's field equations. They also calculated the limit of the mass of a stable relativistic incompressible fluid sphere. After the discovery of neutron stars, a connecting bridge is build up between theoretical and observational astrophysics. Though few studies related to compact star masses \cite{Heap1992,Lattimer2005,Stickland1997,Orosz1999,Van1995}, which are in binaries, have been proposed (with negligible error), there is no information about the radius. Therefore, the theoretical study of the stellar structure is required to follow the correct direction for the newly observed stellar masses to get some new pieces of information. By the compact star modeling some interesting results found in terms of the important parameters like stability factor, radius, compactness, redshift, etc., which cannot be inferred from direct observation. The study of the interior region of the compact stars is not straightforward. Numerous authors \cite{Rahaman2012a,Rahaman2012b,Rahaman2014,Hossein2012,Kalam2012a,Kalam2013,
Kalam2014,Kalam2016,Kalam2017,Kalam2018,
Eur.Phys.J.C.76.5.266,Astrophys.Space.Sci.361.160D,Eur.Phys.J.Plus.129.3,Astrophys.Space.Sci.357.74,Astrophys.Space.Sci.356.2.327,Int.J.Theor.Phys.53.11.3958,Eur.Phys.J.C.77.9.596,Lobo2006,Bronnikov2006,Egeland2007,Dymnikova2002,Chakraborty2017,Singh2021,Sharma2021,Paul2018,Goswami2020,Biswas2020,Aziz2019,Chowdhury2019,Deb2017,Jasim2021} established different models to enlight several features of the interior of the compact objects using various types of metric pleasing Einstein equations as well as all the conditions which are required for a stellar model to be causal.

\par We study the field equations and the mathematical solutions of strange star within Nariai spacetime. Several studies have been performed using the Nariai spacetime by other authors to understand various astrophysical aspects.  Nariai IV metric has also been used to describe neutron stars \cite{Lattimer2001,Lattimer2005,Lattimer2005a} by \cite{Moustakidis2017}. Shaikh et al. \cite{Shaikh2020}  studied the curvature properties of a charged Nariai type spacetime and found that such a metric is not locally symmetric but semi symmetric, and its Ricci tensor is neither Codazzi nor cyclic parallel or recurrent but generalized recurrent. Carlos Batista \cite{Batista2016} presented some solutions which are made of the direct product of several 2-spaces of constant curvature. These solutions turn out to have many magnetic charges, contrary to the usual higher-dimensional generalization of the Nariai spacetime, which has no magnetic charge at all. Fennen and Giulini \cite{Fennen2015} investigated exact solutions to Einstein's equations corresponding to the two spherically symmetric stars, which are made of an incompressible perfect fluid, possibly oppositely charged. The negative pressure of the positive cosmological constant keeps these two stars separated. But there is no cosmological horizon separating these two stars. They showed that the solution gives rise to the Nariai metric or an oversight generalization thereof in the charged case.

\par In this paper, we want to study the physical behavior of the strange star. For this, we have considered the isotropic stellar model to study the fluid sphere. We found a solution of the fluid sphere where metric constants are absolutely depend on star's mass and radius. In this paper we have also discussed the dynamical stability analysis of the system. Our main objective in this study is to give an estimate of the possible radius, central pressure, compactness and redshift of the wide mass range of the strange star.

\par The material of the paper is organised as follows: In Sect. 2 we give the basic field equations in connection to the compact star of Narai IV metric.
In Sect. 3 we have discussed matching condition.
In Sect. 4 we have talk over the physical conditions required of the star model.
In Sect. 5 we review: (5.1) Density and isotropic pressure behavior at the interior of the star, (5.2) Energy conditions, (5.3) TOV equations,
(5.4) Stability condition, (5.5) Compactness and Surface Redshift.
In Sect. 6, we give our concluding remark.

\section{Interior solution}
\label{sec:1}
We consider the static and spherically symmetric metric in isotropic co-ordinates as
\begin{eqnarray}
ds^2 = c^2 e^{\alpha}dt^2-e^{\beta}\left(dr^{2} +r^2d\Omega^{2}\right) \label{eq:1}
\end{eqnarray}
where $\alpha$ and $\beta$ are function of r.
\par Einstein's field equations of gravitation for a non empty space-time are
\begin{eqnarray}\label{eq:2}
R_{ij} - \frac{1}{2} R g_{ij} = -\frac{8\pi G}{c^4}T_{ij}
\end{eqnarray}
where $R_{ij}$ is a Ricci tensor, $T_{ij}$ is energy momentum tensor and R is the scalar curvature.
\par The energy-momentum tensor $T_{ij}$ is defined as
\begin{eqnarray}\label{eq:3}
T_{ij}=\left(p + \rho c^2 \right)v_{i}v_{j} - pg_{ij}
\end{eqnarray}
where $p$ denotes the isotropic pressure and $\rho$ is the density distribution.
\par The velocity vector $v_{i}$ satisfying the relation
\begin{eqnarray}\label{eq:4}
g_{ij}v^{i}v^{j} = 1
\end{eqnarray}
Since the field is static, therefore
\begin{eqnarray}\label{eq:5}
v^{1} = v^{2} = v^{3} = 0 ~~~~~~~~~~~~~~and~~~~~~~~~~~~ v^{4} = \frac{1}{\sqrt{g_{44}}}
\end{eqnarray}
Thus for matter distribution with isotropic pressure the field equation eqn.\ref{eq:2} reduces as \cite{Pant2012}:
\begin{eqnarray}
\frac{8\pi G \rho}{c^2} &=& -e^{-\beta}[\beta^{\prime\prime}+\frac{(\beta^{\prime})^2}{4}+\frac{2\beta^{\prime}}{r}]  \label{eq:6}\\
\frac{8\pi G p}{c^4} &=& e^{-\beta}[\frac{(\beta^{\prime})^2}{4}+\frac{\beta^{\prime}}{r}+\frac{\alpha^{\prime}\beta^{\prime}}{2}+\frac{\alpha^{\prime}}{r}] \label{eq:7}\\
\frac{8\pi G p}{c^4} &=& e^{-\beta}[\frac{\beta^{\prime\prime}}{2}+\frac{\alpha^{\prime\prime}}{2}+\frac{(\alpha^{\prime})^2}{4}+\frac{\beta^{\prime}}{2r}+\frac{\alpha^{\prime}}{2r}] \label{eq:8}
\end{eqnarray}
where, prime($\prime$) denotes differentiation with respect to r. Now from eqn.\ref{eq:7} and eqn.\ref{eq:8} we get the following differential equation in term of $\alpha$ and $\beta$ as
\begin{eqnarray}
\beta^{\prime\prime}+\alpha^{\prime\prime}+\frac{(\alpha^{\prime})^2}{2}-\frac{(\beta^{\prime})^2}{2}-\alpha^{\prime}\beta^{\prime}
-(\frac{\beta^{\prime}}{r}+\frac{\alpha^{\prime}}{r})=0 \label{eq:9}
\end{eqnarray}
Our task is to obtain the physical properties ($\rho$ and $p$) of the fluid sphere from the above equations.
\par We assume that the interior space-time of a star is described by the metric propose by H. Nariai \cite{H. Nariai1950,H. Nariai1951}
\begin{eqnarray}
ds^2 = A\cos^{2}\left(a-\frac{\sqrt 2 M r^{2}}{4}\right)\cos^{-2}\left(b+\frac{ M r^{2}}{4}\right) dt^2\nonumber\\
-A\cos^{-2}\left(b+\frac{ M r^{2}}{4}\right)\left(dr^{2} +r^2d\Omega^{2}\right) \label{eq:10}
\end{eqnarray}
where $A$, $M$, $a$, $b$ are constants.\\
Now using the metric eqn.\ref{eq:10} we get,
\begin{strip}

\begin{equation}\label{eq:11}
8\pi \rho=\frac{M \left(-M r^2 \left(\sin ^2\left(b+\frac{M r^2}{4}\right)+2\right)-6 \sin \left(2 b+\frac{M r^2}{2}\right)\right)}{4 A}
\end{equation}
\begin{footnotesize}
\begin{equation}\label{eq:12}
8 \pi p=\frac{M \cos ^2\left(b+\frac{M r^2}{4}\right) \left(2 \sqrt{2} \tan \left(a-\frac{M r^2}{2 \sqrt{2}}\right) \left(M r^2 \tan \left(b+\frac{M
   r^2}{4}\right)+2\right)+\tan \left(b+\frac{M r^2}{4}\right) \left(3 M r^2 \tan \left(b+\frac{M r^2}{4}\right)+8\right)\right)}{4 A}
\end{equation}
\end{footnotesize}
\end{strip}

\section{Matching Conditions}
It is well known to us that the interior metric of a compact star should be matched to the Schwarzschild exterior solution at the boundary.
\begin{equation}
ds^2 =  \left(1-\frac{2m}{r_{s}}\right)dt^2 -  \left(1-\frac{2m}{r_{s}}\right)^{-1}dr_{s}^2 - r_{s}^2d\Omega^{2} \label{eq:13}
\end{equation}
\par It is always possible to transform this standard form of the Schwarzschild metric into the isotropic form \cite{H. Nariai1950} (see Eq-\ref{eq:1}) by putting
\begin{equation}\label{eq:14}
r_{s}=m+\frac{m^{2}}{4 r}+r
\end{equation}
where $m=$ mass , $r=$ radial co-ordinate.
After using this transformation, the standard Schwarzchild metric transforms into isotropic form as \cite{Hall}
\begin{equation}\label{eq:15}
ds^2 = \left(\frac{1-\frac{m}{2r}}{1+\frac{m}{2r}}\right)^{2} dt^{2}-\left(1+\frac{m}{2r}\right)^{4} \left(dr^{2}+ r^{2}d\Omega^{2}\right)
\end{equation}
The matching of Nariai IV metric at the boundary ($r=R$, where $R$ is radius of the star) with the exterior Schwarzschild metric gives rise to
\begin{small}
\begin{equation}\label{eq:16}
A\cos^{2}\left(a-\frac{\sqrt 2 M R^{2}}{4}\right)\cos^{-2}\left(b+\frac{ M R^{2}}{4}\right)=\left(\frac{1-\frac{m}{2R}}{1+\frac{m}{2R}}\right)^{2}
\end{equation}
\end{small}
\begin{equation}\label{eq:17}
A\cos^{-2}\left(b+\frac{ M R^{2}}{4}\right)=\left(1+\frac{m}{2R}\right)^{4}
\end{equation}

and the continuity of the metric components $g_{tt}$ and $g_{rr}$  at the boundary gives
\begin{eqnarray}\label{eq:18}
M R \left(\sqrt{2} \tan \left(a-\frac{M R^2}{2
   \sqrt{2}}\right)+\tan \left(b+\frac{M R^2}{4}\right)\right)\nonumber\\=-\frac{8m}{4m^{2}-R^{2}}
\end{eqnarray}

\begin{equation}\label{eq:19}
M R \tan \left(b+\frac{M R^2}{4}\right)=-\frac{8m}{2m R+R^{2}}
\end{equation}

We solve the above four eqns.(\ref{eq:16},\ref{eq:17},\ref{eq:18} \& \ref{eq:19}) to get the metric coefficient in terms of the mass (m) and radius (R) of the star as follows

\begin{strip}
\begin{equation}\label{eq:20}
A=\frac{(m-4 R)^2 (m+2 R)^4}{2 (m+4 R) \left(m^5+8 m^4 R+28 m^3 R^2+48 m^2 R^3+40 m R^4+32 R^5\right)}
\end{equation}

\begin{equation}\label{eq:21}
M=\pm \frac{8 \sqrt{2} m (m-4 R)}{(m+2 R) \sqrt{m \left(m^2+6 m R+16 R^2\right) \left(m^3+6 m^2 R+8 m R^2+16 R^3\right)}}
\end{equation}

\begin{eqnarray}\label{eq:22}
b=\mp\frac{2
   \sqrt{2} m R^2 (m-4 R)}{(m+2 R) \sqrt{m \left(m^2+6 m R+16 R^2\right) \left(m^3+6 m^2 R+8 m R^2+16 R^3\right)}}~~~~~~~~~~~~~~~~~~~~~~~~~~~~~~~~~~~~~~~~~~~~~~\\ \nonumber
   +\cos ^{-1}\left(\mp\frac{2 \sqrt{2} R^2 (m-4 R)}{\sqrt{(m+4 R) \left(m^5+8 m^4 R+28 m^3 R^2+48 m^2 R^3+40 m R^4+32 R^5\right)}}\right)
\end{eqnarray}

\begin{equation}\label{eq:23}
a=\pm\frac{4 m R^2 (m-4 R)}{(m+2 R) \sqrt{m \left(m^2+6 m R+16 R^2\right) \left(m^3+6 m^2 R+8 m R^2+16 R^3\right)}}+\cos ^{-1}\left(\pm\frac{4 R^2
   (m-2 R)}{(m+2 R)^3}\right)
\end{equation}
\end{strip}
These two set of solutions also satisfy $p_{|r=R}=0$ equation i.e. pressure get vanished at the boundary.

\section{Conditions for well behaved solution}
For well-behaved nature of the solution in spherical coordinates, the following conditions should be satisfied \cite{Delgaty1998},\cite{Pant2010,Pant2012,PantA2012}:
\begin{enumerate}
\item[(i)] The solution should be free from physical and geometrical singularities i.e. finite and positive values of central density and central pressure i.e. $\rho_{c} > 0$ and $p_{c} > 0$.
\item[(ii)] The solution should give rise to positive and monotonically decreasing expressions for density($\rho$) and pressure(p) with the increase of r.
\item[(iii)] The pressure at the boundary should vanish.
\item[(iv)] The ratio of pressure to the density (p/$\rho$) should be monotonically decreasing with the increase of r. $(\frac{d\rho}{dr})_{r=0} = 0$ and $(\frac{d^{2}\rho}{dr^{2}})_{r=0} < 0$ so that density gradient $\frac{d\rho}{dr}$ is negative for $0 < r \leq R$.  $(\frac{dp}{dr})_{r=0} = 0$ and $(\frac{d^{2}p}{dr^{2}})_{r=0} < 0$ so that pressure gradient $(\frac{dp}{dr})$ is negative for $0 < r \leq R$. These above conditions imply that pressure and density should be maximum at the centre and monotonically decreasing towards the pressure-free interface.
\item[(v)] The solution should satisfy all the energy conditions i.e. weak energy condition, strong energy condition, null energy condition and dominant energy condition.
\item[(vi)] The solution should obey the casualty condition $v_s^2=(\frac{dp}{d\rho})\leq 1$ i.e. speed of sound should be less than that of light throughout the model.
\item[(vii)] Stellar equilibrium exist everywhere within the star.
\item[(viii)]  When the observer and the source of light are in relative motion, the observer, instead of receiving the light at its original wavelength, receives the light of a slightly shifted wavelength. The shift causes an increase in the wavelength and a decrease in the frequency or energy of electromagnetic radiation called redshift. The redshift is written as follow
\begin{equation} \label{eq:38a}
Z(r)=\frac{1}{\sqrt{1-\frac{2M(r)}{r})}}-1
\end{equation}. 

The redshift ($Z$) should be positive and finite for the stellar model.The maximum allowed value of surface redshift is $Z\leq 0.85$ \cite{Haensel2000}.
\end{enumerate}
Under these above conditions, the solution should be well behaved.

\section{Analysis of Physical properties}

\subsection{Density and Pressure Behavior of the star}
The internal metric is free from physical and geometrical singularities. Thus central density and central pressure must be positive and finite i.e. $\rho_{c} > 0$ and $p_{c} > 0$. The central density and pressure are given by
\begin{equation}\label{eq:24}
\rho_{c}=-\frac{3 M \sin (b) \cos (b)}{8 \pi  A}>0
\end{equation}

\begin{equation}\label{eq:25}
p_{c}=\frac{M \cos ^2(b) \left(\sqrt{2} \tan (a)+2 \tan (b)\right)}{8
   \pi  A}>0
\end{equation}
and since the energy density and pressure is monotonically decreasing function, we have
\begin{eqnarray}\label{eq:26}
\frac{d \rho}{d r}\bigg|_{r=0} = 0 ~~~~~~~~~~~~~~and~~~~~~~~~~~~ \frac{d p}{d r}\bigg|_{r=0} = 0
\end{eqnarray}
and
\begin{eqnarray}
 \frac{d^{2}\rho}{dr^{2}}\bigg|_{r=0} &=&-\frac{5 M^2 \cos ^2(b)}{16 \pi  A} <0 \label{eq:27}\\
  \frac{d^2p}{dr^{2}}\bigg|_{r=0} &=&\frac{M^2 \left(2-2 \sec ^2(a) \cos ^2(b)-\sin ^2(b)\right)}{16 \pi  A} <0 \label{eq:28}\nonumber \\
  \end{eqnarray}

  \begin{figure}[!ht]
\includegraphics[scale=0.59]{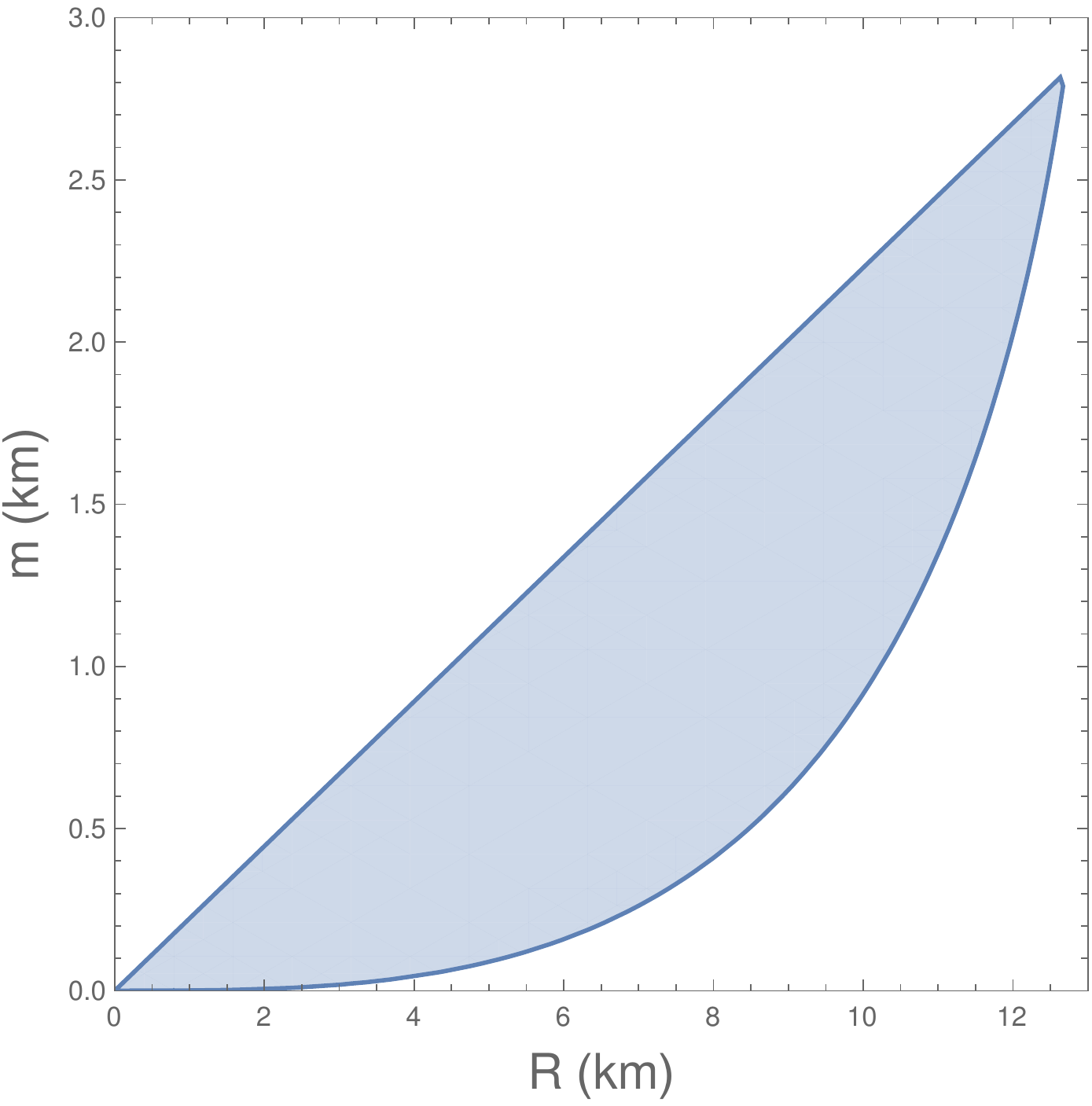}
\caption{Graphical presentation of the allowed the mass-radius region}
\label{fig1}
\end{figure}

\begin{figure}[!ht]
\includegraphics[scale=0.295]{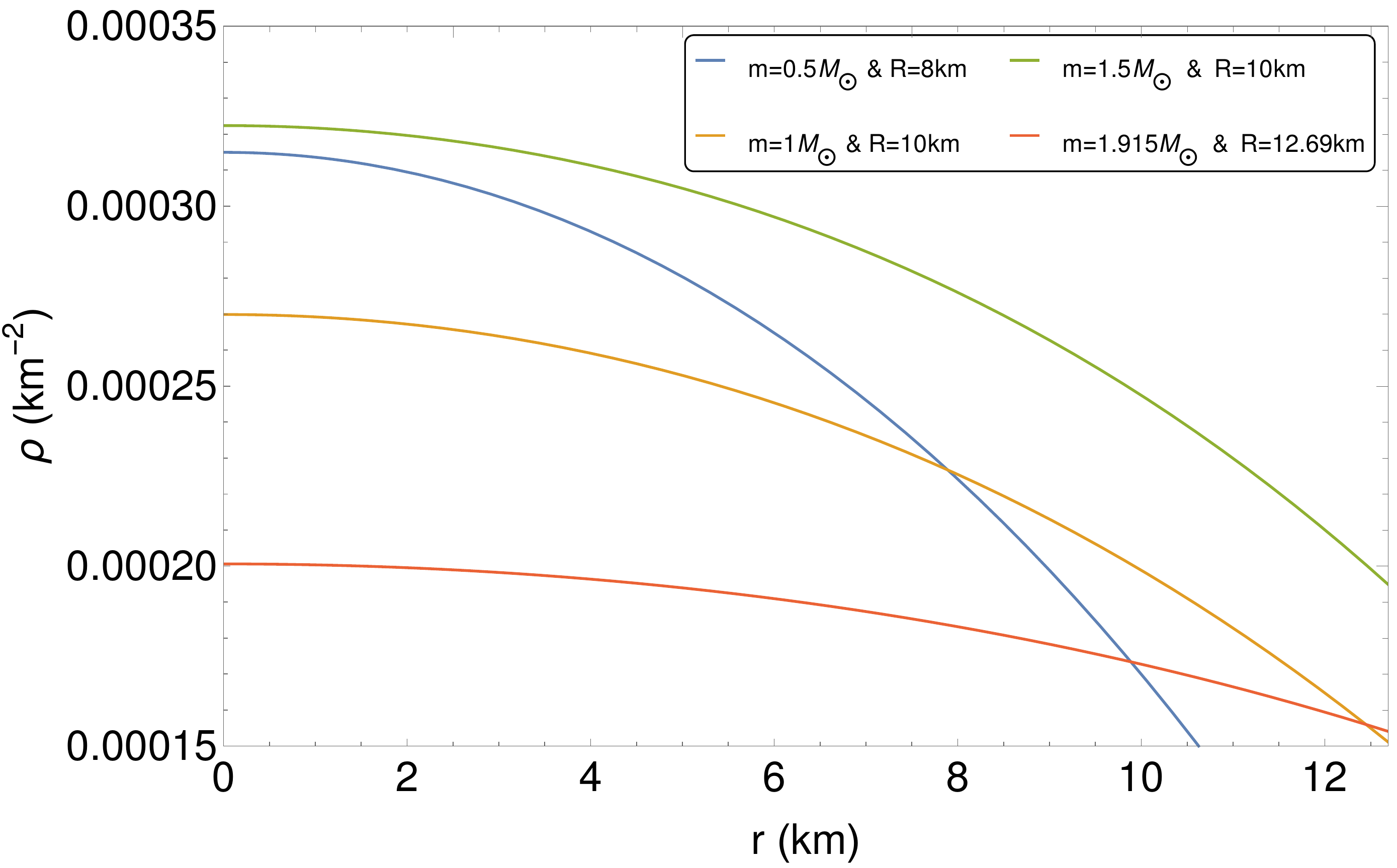}
\caption{Variation of the density $\rho$ against radial parameter $r$ of the star}
\label{fig2}
\end{figure}

\begin{figure}[!ht]
\includegraphics[scale=0.295]{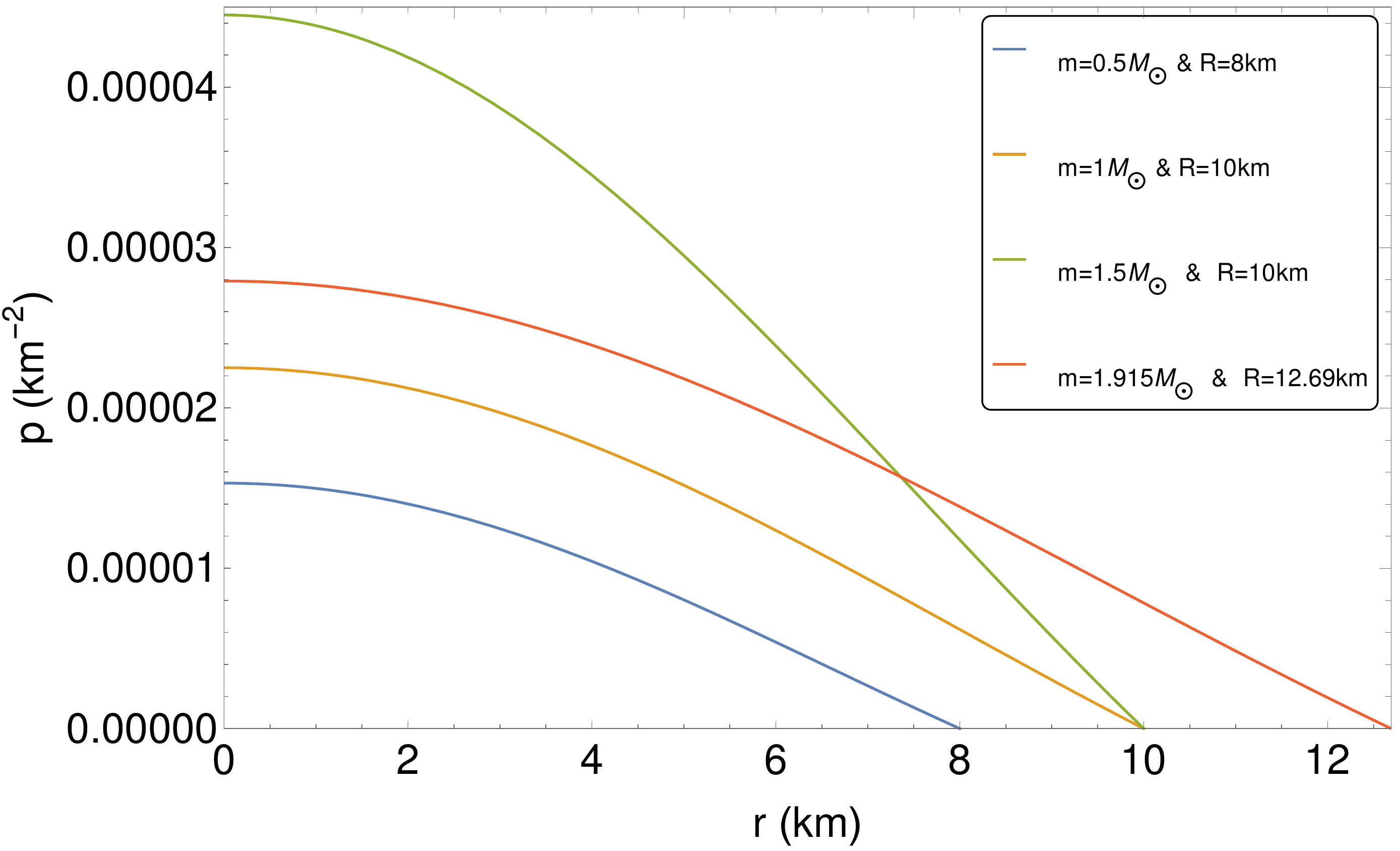}
\caption{Variation of the pressure $p$ against radial parameter $r$.}
\label{fig3}
\end{figure}

\begin{figure*}[!ht]
	\includegraphics[scale=0.45]{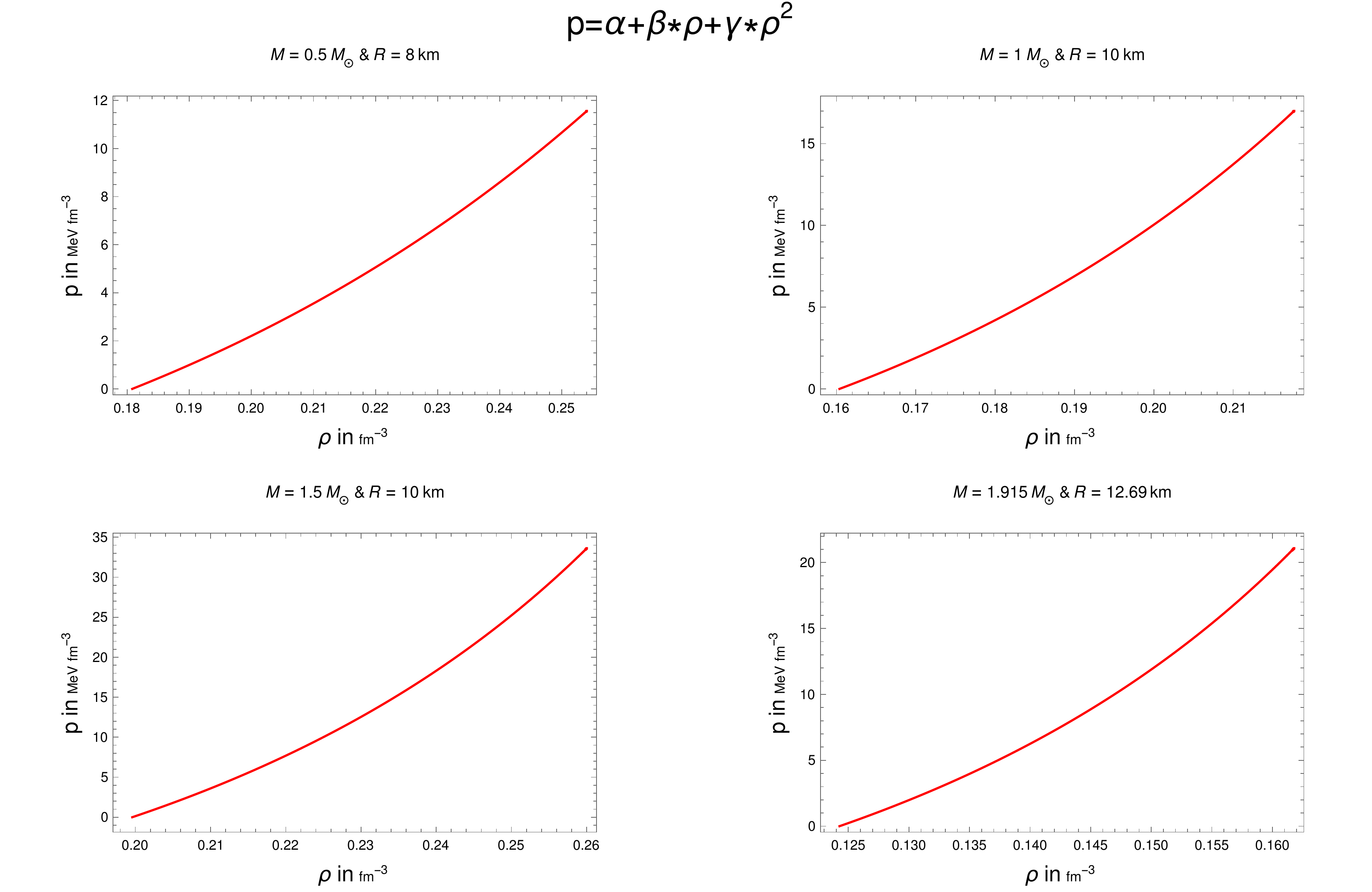}
	\caption{Equation of state of four strange stars, where $ \alpha, \beta, \gamma $ are constants.}
\end{figure*}

\begin{figure*}[!ht]
\includegraphics[scale=0.45]{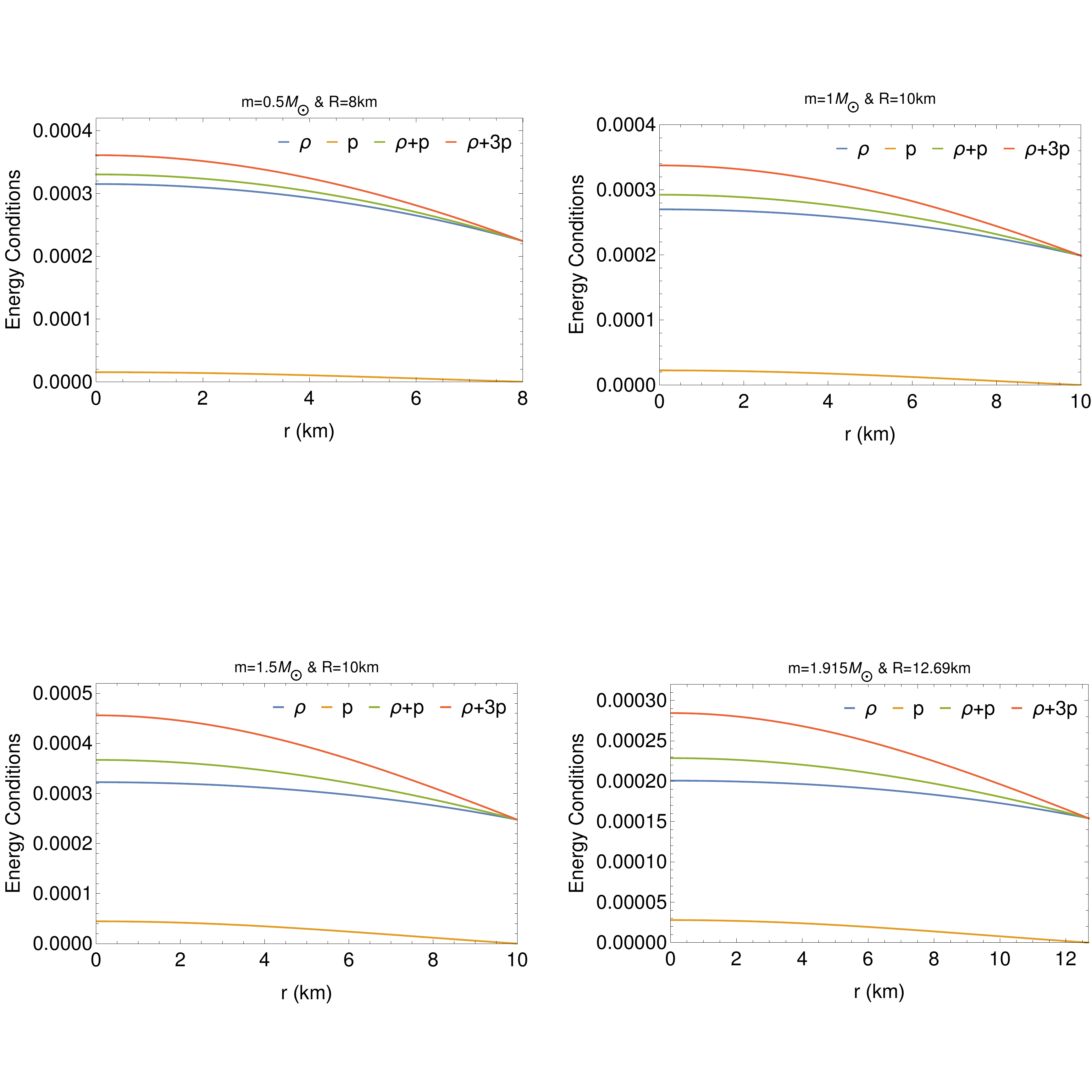}
\caption{Energy conditions at the stellar interior}
\label{fig4}
\end{figure*}

\begin{figure*}[!ht]
\includegraphics[scale=0.45]{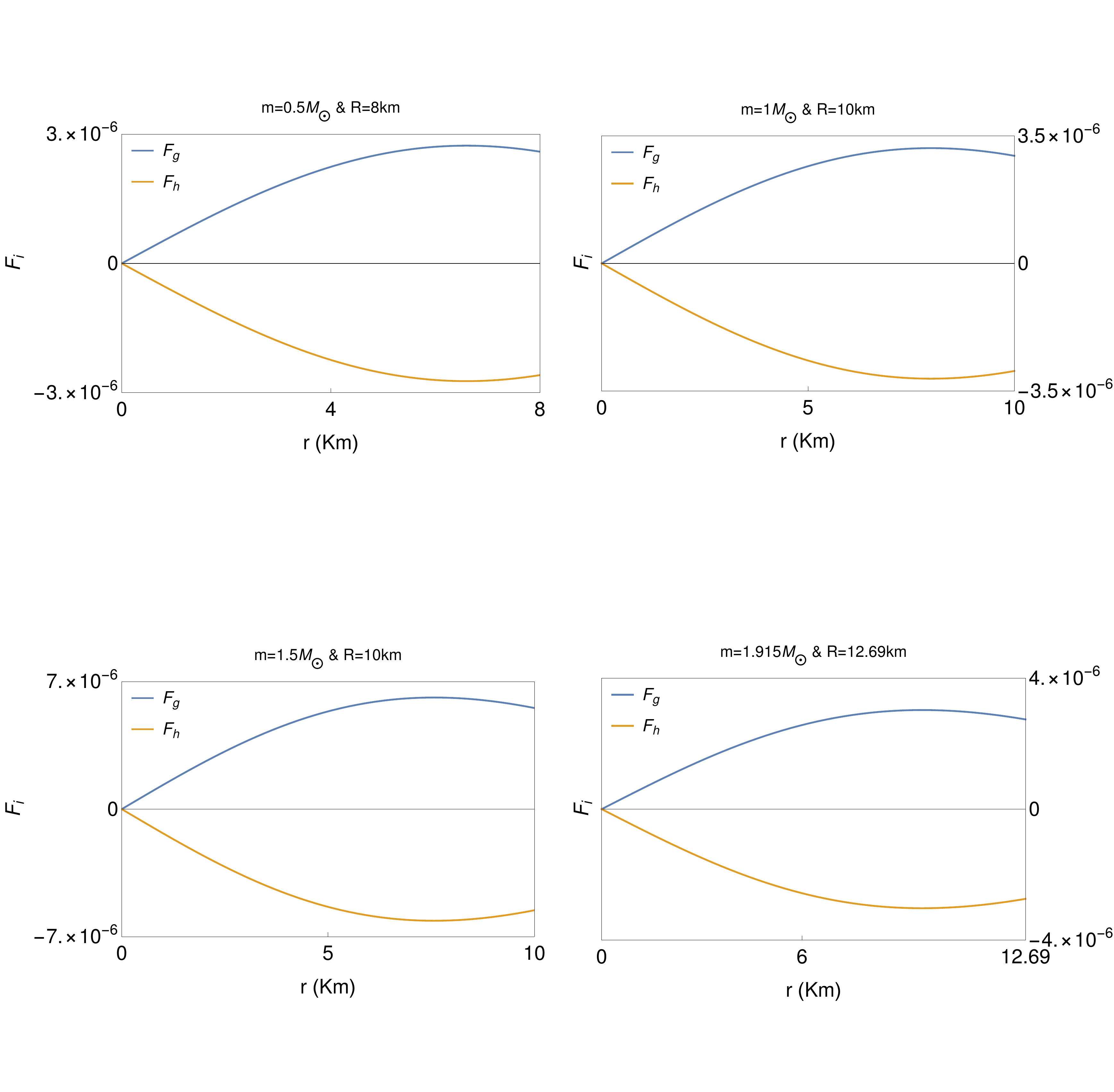}
\caption{Behaviors of gravitational and hydrostatic force at the stellar interior of the star}
\label{fig5}
\end{figure*}

\begin{figure}[!ht]
\includegraphics[scale=0.298]{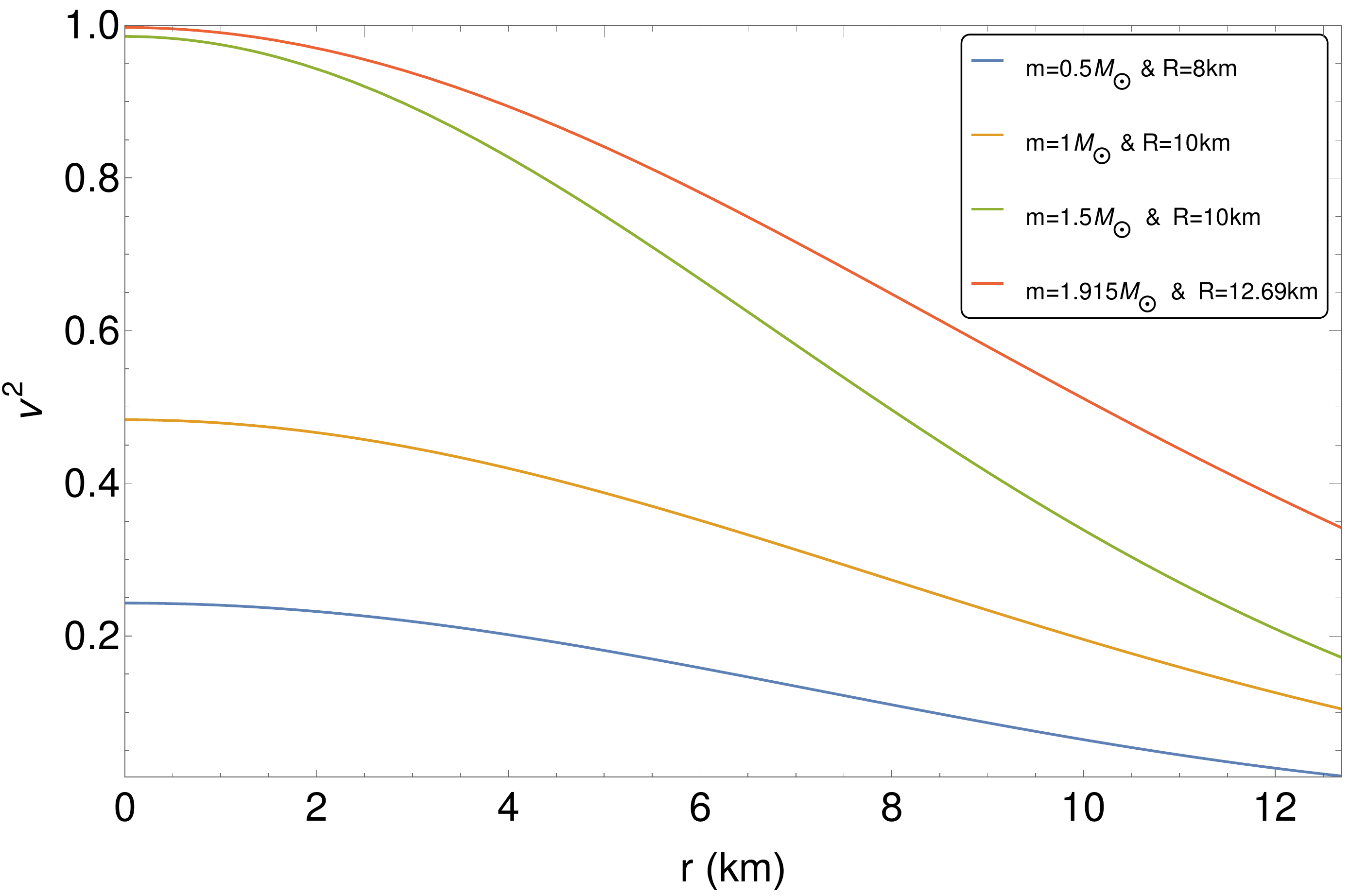}
\caption{Variation of the sound speed at the stellar interior.}
\label{fig8}
\end{figure}

\begin{figure}[!ht]
\includegraphics[scale=0.298]{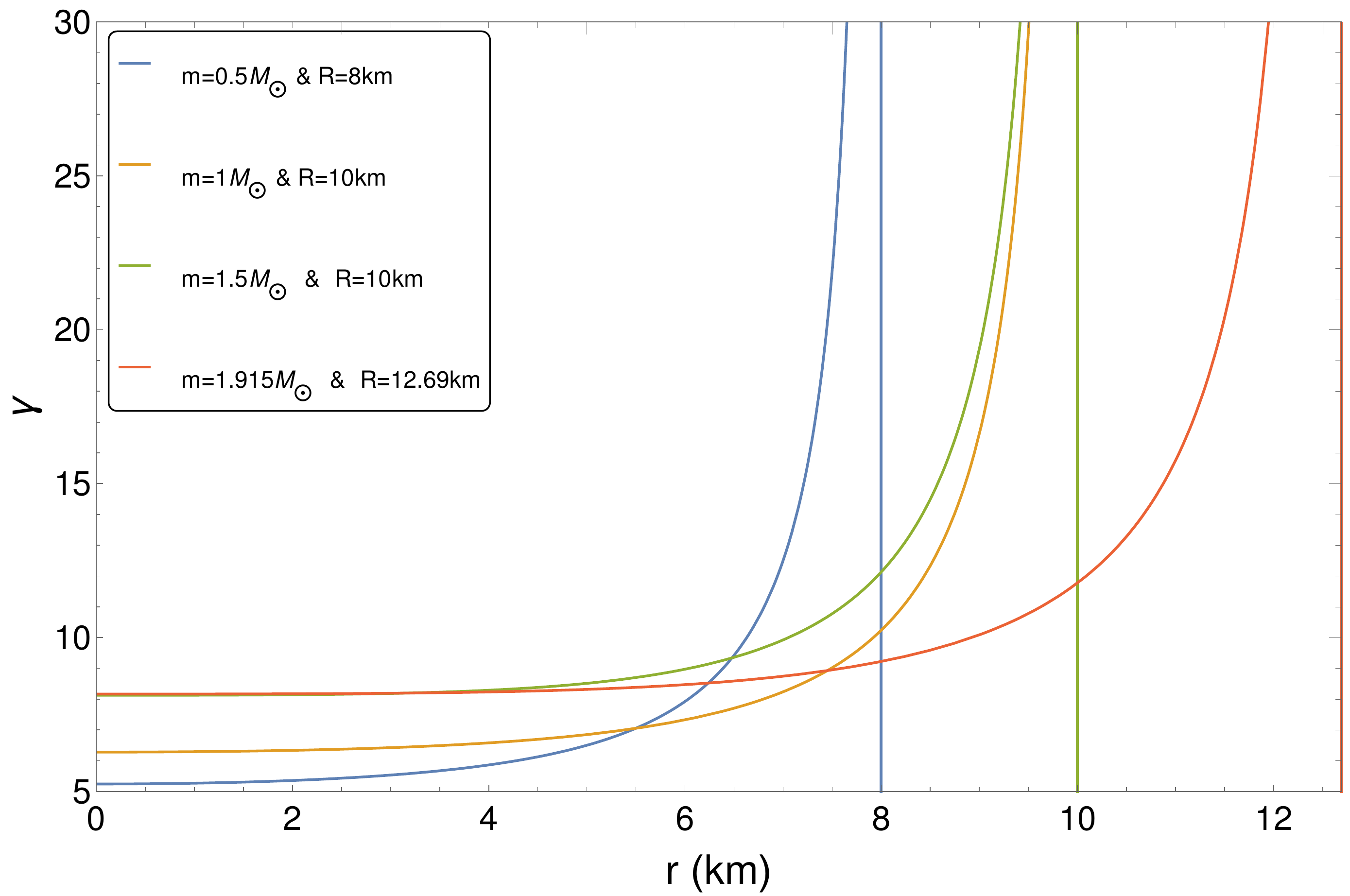}
\caption{ Variation of the adiabatic index at the stellar interior.}
\label{fig9}
\end{figure}

\begin{figure}[!ht]
\includegraphics[scale=0.265]{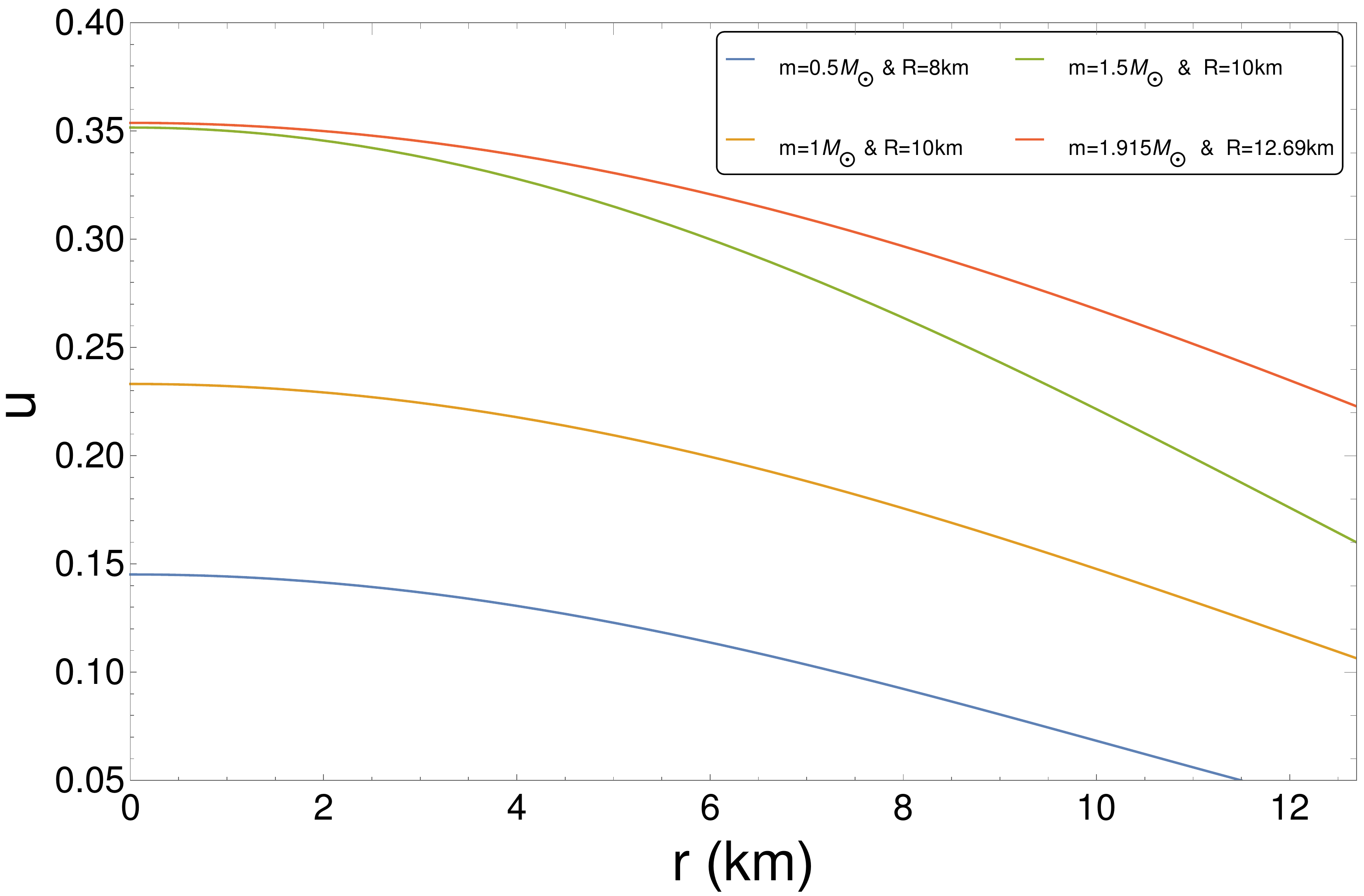}
\caption{Variation of the compactness $u$ against radial parameter $r$.}
\label{fig6}
\end{figure}

\begin{figure}[!ht]
\includegraphics[scale=0.298]{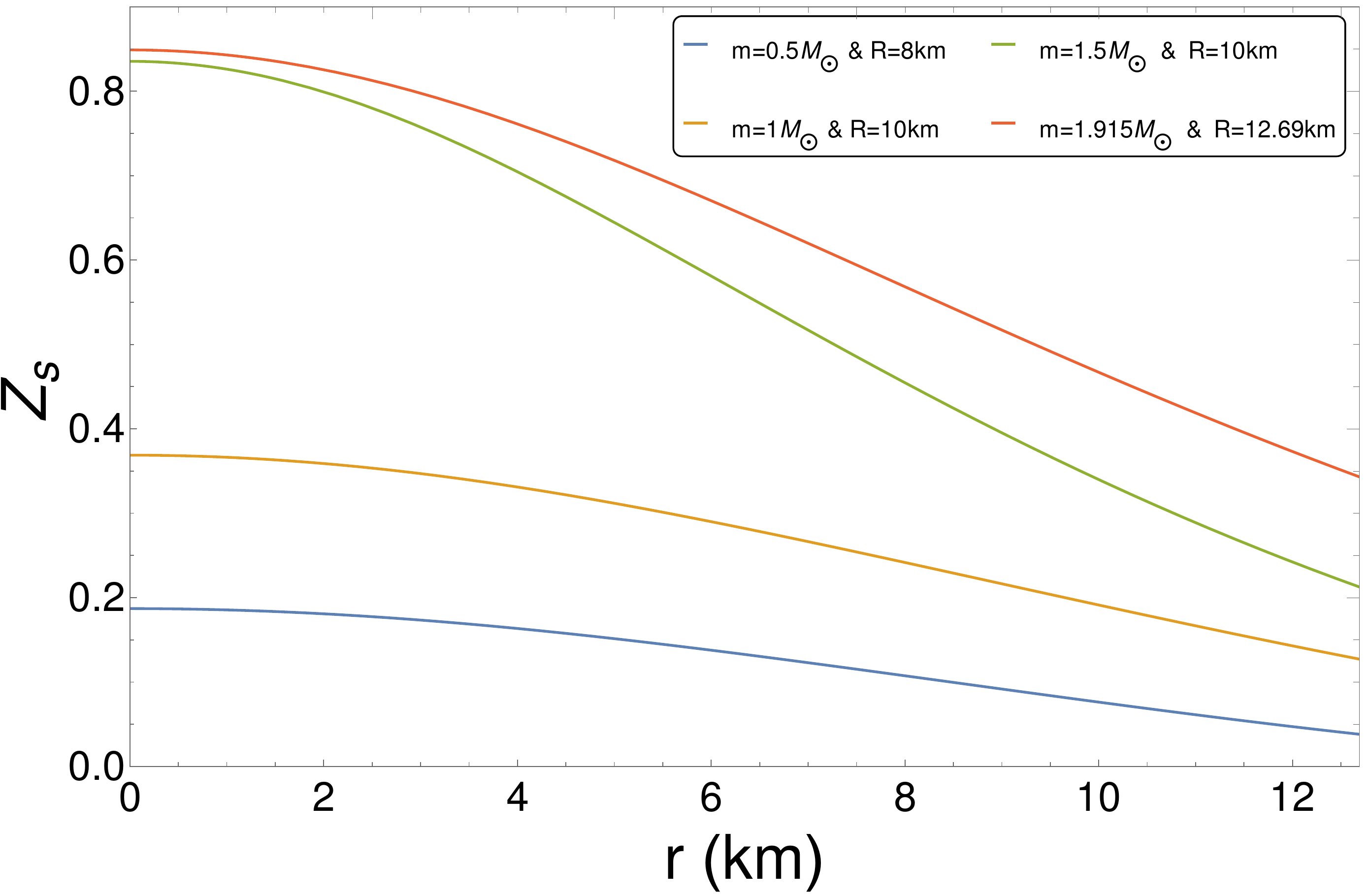}
\caption{Variation of the redshift $Z_{s}$ against radial parameter $r$ of the star}
\label{fig7}
\end{figure}

\subsection{Energy conditions}
From Fig.~5, we see that all the energy conditions such as null energy condition (NEC), weak energy condition (WEC), strong energy condition (SEC) and dominant energy condition (DEC) satisfies at every point in the interior of the compact star simultaneously. These energy conditions are as follows:
 \\
\begin{eqnarray}\label{eq:29}
\mbox{i)  NEC:}~~~~~~~~~~~~~~~~~~~~~~~~~~\rho+p \geq 0\nonumber \\
 \mbox{ii) WEC:}~~~~~~~~~~~~~~~~~ \rho+p  \geq  0  , \rho  \geq  0  \nonumber \\
\mbox{iii) SEC:}~~~~~~~~  \rho+p \geq  0 , \rho+3p  \geq  0 \nonumber \\
    \mbox{iv) DEC:}~~~~~~~~~~~~~~~~~~~~~~~~~~~~~ \rho >|p|
\end{eqnarray}

\subsection{TOV equation}
The stellar equation (some authors considered it as TOV equation) can be written as
\begin{equation}\label{eq:30}
\frac{dp}{dr} +\frac{1}{2} \nu^\prime\left(\rho
 + p\right)
= 0
\end{equation}
The stellar equation (TOV equation) describes the equilibrium condition for the strange star subject to the effective gravitational ($F_g$) and effective hydrostatic ($F_h$) force nature of the stellar object as
\begin{equation}
F_h+ F_g  = 0,\label{eq:31}
\end{equation}
where,
\begin{eqnarray}
F_g &=& \frac{1}{2} \nu^\prime\left(\rho+p\right)\label{eq:32}\\
F_h &=& \frac{dp}{dr} \label{eq:33}
\end{eqnarray}
Therefore, from Fig.~6 we see that the static equilibrium do exist in the presence of gravitational and hydrostatic forces.

\subsection{Stability}
For a physically acceptable model, speed of sound should be within the range
$0 \leq  v^2=(\frac{dp}{d\rho})\leq 1$  \cite{Herrera1992,Abreu2007,Chengjun Xia2021, Ma2019,Bedaque2015,Cherman2009,Hohler2009,Benincasa2006,Mateos2007,Benincasa2006a,Borsanyi2010,Alford2013}. Fig-\ref{fig8} indicates that the sound speed is monotonically decreasing function of the radial distance. 
The sound speed at the center is given by
\begin{equation}\label{eq:34}
0 \leq v_{c}^2=\frac{1}{10} \left(4 \sec ^2(a)+2 \tan ^2(b)-4 \sec ^2(b)\right)\leq 1
\end{equation}
The adiabatic Index should be \cite{Chandrasekhar1964},
\begin{equation}
\gamma=\frac{p+\rho}{p}\frac{dp}{d\rho} >\frac{4}{3}
\end{equation}
Fig-\ref{fig9} shows that the adiabatic index within the star is always greater than $\frac{4}{3}$.
The adiabatic index at the centre is give by
\begin{small}
\begin{equation}\label{eq:36}
\gamma_{c}=\frac{\left(\tan (b)-\sqrt{2} \tan (a)\right) \left(-2 \sec ^2(a)+\sec ^2(b)+1\right)}{5 \left(\sqrt{2} \tan (a)+2 \tan (b)\right)}>\frac{4}{3}
\end{equation}
\end{small}
\subsection{Compactness and Surface Redshift}
From eqn.\ref{eq:17}, we get the compactness and redshift written as follow
\begin{equation}\label{eq:37}
u(r)=\frac{m(r)}{r}=2 \left(\sqrt[4]{A \sec ^2\left(b+\frac{M r^2}{4}\right)}-1\right)
\end{equation}
and
\begin{equation} \label{eq:38}
Z(r)=\frac{1}{\sqrt{1-4 \left(\sqrt[4]{A \sec ^2\left(b+\frac{M r^2}{4}\right)}-1\right)}}-1
\end{equation}
According to Buchdahl \cite{Buchdahl1959}, for a static spherically symmetric perfect fluid allowable mass-radius ratio should be $\frac{2M}{R} <\frac{8}{9}$ and Mak \cite{Mak2001} also gave a more generalized expression for the same manner. The maximum allowed value of surface redshift is $Z\leq 0.85$ \cite{Haensel2000}. The central compactness and the central redshift are given by
\begin{equation}\label{eq:39}
u_{c}=2 \left(\sqrt[4]{A \sec ^2(b)}-1\right)<\frac{4}{9}
\end{equation}
\begin{equation}\label{eq:40}
 Z_{c}=\frac{1}{\sqrt{1-4 \left(\sqrt[4]{A \sec ^2(b)}-1\right)}}-1 \leq 0.85
\end{equation}
Figs. \ref{fig6} \& \ref{fig7} show that the compactness and redshift within the star are always less than $\frac{4}{9}$ and less than to the value of $0.85$.
\par Now if we solve all these conditions \\ (Eqs-\ref{eq:24},\ref{eq:25},\ref{eq:26},\ref{eq:27},\ref{eq:28},\ref{eq:34},\ref{eq:36},\ref{eq:39},\ref{eq:40}) alongside the demand that the central density of the star \cite{Bombaci2018} has at least of the order of the nuclear density i.e. $2.7\times 10^{14} g/cc$, then we get the important mass-radius relation (Fig-\ref{fig1}). Fig-\ref{fig1} presents the allowed region of the mass-radius graph. The minimum radius for a given mass ($m$) is $R_{min}=4.4857m $ km ($m$ in $km$). Table-\ref{table:1} enlists the maximum radius $R_{max}$, minimum central pressure $p_{c(max)}$, minimum compactness $u_{min}$ and minimum redshift $Z_{min}$ for a given mass. We see from the Figs-\ref{fig2}-\ref{fig3} that the energy density and the pressure are monotonically decrease with the radial distance.   

\begin{table*}[!ht]
\caption{List of the maximum radius for a given mass of a strange star, and the corresponding minimum central pressure, minimum compactness and minimum redshift}
\label{table:1}
\begin{tabular}{@{}|p{2cm}|p{3cm}|p{4cm}|p{3cm}|p{3cm}|@{}}
\hline
Mass ($M_{\odot}$) & Maximum Radius ($R_{max}$ in km)&Minimum Central pressure ($p_{c(min)}$ in $10^{-6}km^{-2}$) & Minimum Compactness ($u_{min}$)& Minimum Redshift ($Z_{min}$) \\[1ex]
\hline
 0.01 & 2.78414  & 0.50505 & 0.005304& 0.005346 \\
 0.05 & 4.70399  & 1.51260 & 0.015695 & 0.016075 \\
 0.1 & 5.86247  & 2.45425 & 0.025187 & 0.026181 \\
 0.2 & 7.25713  & 4.03779 & 0.040694 & 0.043359 \\
 0.3 & 8.18254  & 5.45781 & 0.054137 & 0.058972 \\
 0.4 & 8.88272  & 6.80319 & 0.066493 & 0.073957 \\
 0.5 & 9.44583 & 8.11042 & 0.078161 & 0.088709 \\
 0.7 & 10.3150  & 10.6820 & 0.100205 & 0.118320 \\
 0.8 & 10.6616  & 11.9675 & 0.110797 & 0.133436 \\
 0.9 & 10.9653  & 13.2622 & 0.121195 & 0.148886 \\
 1. & 11.2366  & 14.5567 & 0.131409 & 0.164696 \\
 1.2 & 11.6854 & 17.2502 & 0.151634 & 0.198029 \\
 1.3 & 11.8765  & 18.6274 & 0.161627 & 0.215590 \\
 1.4 & 12.0482  & 20.0345 & 0.171580 & 0.233871 \\
 1.5 & 12.2027  & 21.4746 & 0.181508 & 0.252957 \\
 1.6 & 12.3417  & 22.9511 & 0.191428 & 0.272937 \\
 1.7 & 12.4668  & 24.4672 & 0.201352 & 0.293914 \\
 1.8 & 12.5791  & 26.0261 & 0.211292 & 0.315999 \\
 1.9 & 12.6799  & 27.6314 & 0.221258 & 0.339318 \\
 1.91 & 12.6894  & 27.7946 & 0.222256 & 0.341723 \\
 1.915 & 12.6941  & 27.8764 & 0.222755 & 0.342930 \\
 1.916 & 12.6950  & 27.8927 & 0.222855 & 0.343172 \\
 1.91675 & 12.6957  & 27.9050 & 0.222930 & 0.343354 \\
\hline
\end{tabular}
\end{table*}

\begin{table*}[!ht]
\caption{List of the maximum value of some parameters corresponding to minimum radii $R_{min}=4.4857m $ km ($m$ in $km$)}
\label{table:2}
\begin{tabular}{@{}|p{3cm}|p{3cm}|p{3cm}|p{3cm}|p{3cm}@{}|}
\hline
Maximum Central Energy Density ($\rho _{c(max)}$) in $km^{-2}$ & Maximum Surface Energy Density ($\rho _{R(max)}$) in $km^{-2}$ & Maximum Central Pressure ($p_{c(max)}$) in $km^{-2}$ & Maximum Compactness ($u_{max}$)& Maximum Redshift ($Z_{max}$) \\[1ex]
\hline
& & & & \\
 $\frac{0.00160562}{m^2}$ & $\frac{0.0012333}{m^2}$ & $\frac{0.000223533}{m^2}$ &$0.222841$&$0.343188$ \\[1ex]
 \hline
\end{tabular}
\end{table*}

\section{Conclusions}
In this article, we present a strange star model(having mass up to $1.9165M_{\odot}(=2.81 km)$) by using the Nariai IV metric. Using this model, we study the strange stars physical properties. We present a mass-radius region where all the regularity conditions, energy conditions, the TOV equation, and stability conditions are satisfied.  We show, in Fig. 4, how the pressure depends on the energy in four cases; and found that the mathematical relation between them has a form like $ p=\alpha+\beta \rho+\gamma \rho^2 $. From the EOS graph (see Fig. 4), one can see that the slope of these plots is high, which indicates that our EOS is  stiff. This kind of stiffness in EOS is found in SQM(1-3) EOS \cite{Prakash1995,Lattimer2001}. The maximum mass ($1.9165M_{\odot}$) found in our model is also very similar to the maximum mass (around $2M_{\odot}$) found for SQM3 EOS (see Fig.2 of \cite{Lattimer2001}). That's why we are claiming that strange stars are made of, mainly, quarks with a little bit mixture of baryons.  According to the Ref. \cite{Chengjun Xia2021}, we see that our model fulfil the causality limit ( $ \frac{c_s}{c} < 1 $ ). We see in the Ref. \cite{Chakraborty2017} that for hybrid star ( which is made by quark matter and nuclear(baryonic) matter, considered as two different fluids) square of sound speed is less than 1/3. We also see in this Ref. \cite{Alford2013} that the square of sound speed may be greater than 1/3 when the quark matter is strongly coupled. Therefore, we can argue that our strange star model is applicable for both cases: (i) strange star made by quark matter which is strongly coupled in which square of sound speed greater than 1/3 (ii) Also hybrid star made by quark matter and nuclear(baryonic) matter in which square of sound speed less than 1/3. According to our model,   strange stars having a mass greater than $1.9165M_{\odot}$ violates stability conditions. This upper limit of the mass of a strange star comes out as a consequence of imposing regularity conditions, stability conditions and taking the minimum central density of the value of nuclear density ($2.7\times 10^{14} g/cc$) \cite{Bombaci2018}. For a given mass ($m$), we get a range of radius by imposing the regularity conditions, energy conditions, the TOV equation, and stability conditions. The minimum radius for a given mass ($m$) of a strange star is $R_{min}=4.4857m $ km ($m$ in $km$) , and the upper limit of the mass is enlisted in Table-\ref{table:1}. This maximum radius is the immediate result of imposing the fact that the central density is at least of the order of nuclear density. If we change the minimum central density from the order of the nuclear density, then the maximum radius for a given mass also deviates from the Table-\ref{table:1}. We see that as we increase the minimum central density (here, we take the minimum central density as $2.7\times 10^{14} g/cc$ ), the upper bound of radius for a given mass get squeezed. However, the minimum radius of that strange star does not depend on the minimum central density at all. It is determined by the fact that causality condition must be obeyed. We substitute the value of $R_{min}$ in eqn.-\ref{eq:24} \& \ref{eq:25} to get the maximum value of the central energy density and the central pressure, and we get the value of maximum surface energy density, surface compactness and surface redshift by substituting the value of $R_{min}$ in eqn.-\ref{eq:11},\ref{eq:37} \& \ref{eq:38} respectively which is depicted in Table-\ref{table:2}. Here we see interesting results. The maximum central energy density and pressure and the maximum surface energy density are inversely proportional to the square of the mass of a strange star corresponding to the minimum radius. But the maximum compactness and the maximum redshift is independent of the mass of the strange star. For a given mass, we get a range of the central energy density as well as a range of the radius of the star from our model. Therefore, given the mass, one can get the radius by fixing the central density or vice versa. As we approach the maximum mass, the range of radius for a given mass gets shrink down. For greater than one solar mass, this range of radius is very narrow, which is useful to predict the radius of a strange star from an only input parameter that is the mass of the strange star. Therefore, we think that this model can be very beneficial to predict the radius for a strange star of mass greater than one solar mass.

\footnotesize
\section*{Acknowledgments}
MM is thankful to CSIR (Grand No.-09/1157(0007)/ 2019-EMR-I) for providing financial support. SM and MK is grateful to the Inter-University Centre for Astronomy and Astrophysics (IUCAA), Pune, India for providing Associateship programme under which a part of this work was carried out.

\end{document}